\documentclass[aps,showpacs,twocolumn,floatfix,bibnotes]{revtex4}

\usepackage{here}

\usepackage[dvips]{graphicx}

\newcommand\pictc[5]{\begin{figure}
                       \centerline{
                       \includegraphics[width=#1\columnwidth]{#3}}
                   \protect\caption{\protect\label{fig:#4} #5}
                    \end{figure}            }
\newcommand\pict[4][1.]{\pictc{#1}{!tb}{#2}{#3}{#4}}
\newcommand\rpict[1]{\ref{fig:#1}}

\newcommand\leqt[1]{\protect\label{eq:#1}}
\newcommand\reqtn[1]{\ref{eq:#1}}
\newcommand\reqt[1]{(\reqtn{#1})}

\newcounter{Fig}

\begin{document}
\begin{sloppy}

\title{Beam shaping by a layered structure with left-handed materials}

\author{Ilya V. Shadrivov}
\author{Andrey A. Sukhorukov}
\author{Yuri S. Kivshar}

\affiliation{Nonlinear Physics Group, Research School of Physical
Sciences and Engineering, Australian National University, Canberra
ACT 0200, Australia}
\homepage{http://www.rsphysse.anu.edu.au/nonlinear}

\begin{abstract}
We analyze transmission of a layered photonic structure (a one-dimensional photonic crystal) consisting of alternating slabs of two materials with positive and negative refractive index. For the periodic structure with {\em zero
averaged refractive index}, we demonstrate a number of unique
properties of the beam transmission observed in strong beam
modification and reshaping.
\end{abstract}


\maketitle

Recent experimental results confirmed the existence of a novel
type of composite materials, the so-called {\em left-handed (LH)
metamaterials} (or materials with both negative permittivity and
permeability), which possess the property of {\em negative
refraction}. Such materials were suggested theoretically a long time
ago \cite{veselago}, but they have recently attracted much attention due to
their experimental realization and recent `hot' debates on the use
of a LH slab as a perfect lens to focus both propagating and
evanescent waves \cite{nature}. The perfect lens concept was first
suggested by Pendry \cite{Pendry:2000-3966:PRL}, who demonstrated
that a slab of a lossless negative-refraction material can provide
a perfect image of a point source. Although the perfect image is a
result of an ideal theoretical model employed in the analysis of
Ref. \cite{Pendry:2000-3966:PRL}, the resolution limit was shown
to be independent on the wavelength of electromagnetic wave (but
can be determined by other sources such as loss, spatial
dispersion, etc.),  and it can be better than the resolution of a
conventional lens \cite{Luo:2002-201104:PRB}.

Multilayered structures containing LH materials can be considered
as a sequence of the perfect lenses and, therefore, they are expected to
possess novel and unique transmission properties.
Such multilayered structures have been investigated theoretically
for calculating the transmittance or reflectance in the Bragg
regime \cite{zhang,gerardin}. More recently, it was shown that a
one-dimensional stack of layers with alternating right- and
left-handed materials with zero averaged refractive index displays
a narrow spectral gap in the transmission
\cite{Chan:DownUnder,WuHe:arXiv}, which is quite different from a Bragg reflection gap.

In this Letter, we study the properties of layered photonic
structures consisting of alternating slabs of positive and
negative refractive index materials (see Fig. \rpict{geom}), and demonstrate unusual angular dependencies for the transmission of such slabs when the averaged refractive index is close to zero. We demonstrate how these properties can be employed for strong beam reshaping.

\pict{fig01.eps}{geom}{Schematic of a multilayered structure consisting
of slabs with alternating right- and left-handed materials.}

We consider a one-dimensional photonic crystal formed by
alternative slabs, as schematically shown in Fig. \rpict{geom}. We assume that the periodic structure is made from the materials of two types: the slabs of the width $b$ are made of a LH material, and they are separated by
the right-handed (RH) layers of the width $a$. Such a structure
can be treated as an array of Pendry's perfect lenses with the
variation of the refractive index in one structural period,
\begin{equation} \leqt{Index}
    n(z)= \left\{
    \begin{array}{lr}
        n_r = \sqrt{ \epsilon_r \mu_r } , \; 0<z<a;  \\
        n_l = -\sqrt{ \epsilon_l \mu_l }, \; a<z<a+b=\Lambda,
    \end{array}
    \right.
\end{equation}
where $n_r$ and $n_l$ are the positive and negative indices of
refraction of the RH and LH materials, respectively. First, we
consider waves propagating in the ($x,z$) plane with the wave
vectors ${\bf k}=(k_x,0,k_z)$, which are TE-polarized, i.e. they
have the only component $E=E_y$ described by the
Helmholtz-type equation,
\begin{equation} \leqt{Helm}
   \Delta E 
   + \frac{\omega^2}{c^2} n^2(z) E  
   - \frac{1}{\mu(x)} \frac{\partial \mu}{\partial x} E
   = 0,          
\end{equation}
where $\Delta$ is the two-dimensional Laplacian. In an infinite
periodic structure the propagating waves have the form of {\em Bloch
modes}, which electric field envelopes satisfy the
periodicity property, $E(z+\Lambda) = E(z) \exp(i K_b)$. Here
$K_b$ is the normalized Bloch wave number, and it can be found as a solution of the standard equation for two-layered periodic structures (see, e.g., Ref.~\cite{WuHe:arXiv}),
\begin{eqnarray} \leqt{TransferM}
   2 \cos(K_b) = 2 \cos{ \left( k_{rz} a + k_{lz} b \right)} -\nonumber\\
   - \left( \frac{p_r}{p_l} + \ \frac{p_l}{p_r} - 2 \right)
            \cdot \sin{\left( k_{rz} a  \right)}
            \cdot \sin{\left( k_{lz} b  \right)}.
\end{eqnarray}
Here $k_{rz}$ and $k_{lz}$ are the $z$-components of the
wavevector in the RH and LH media, respectively, and $p_{r,l} =
\sqrt{\epsilon_{r,l}/\mu_{r,l}}, \: cos(\theta_{r,l})$, where
$\theta_{r,l} = tan^{-1}\left( k_{(r,l) x}/k_{(r,l) z} \right) $ are the propagation angles in the corresponding media. For real $K_b$ the Bloch waves are propagating; complex $K_b$ indicate the presence of {\em band gaps}, where the wave propagation is inhibited.

\pict{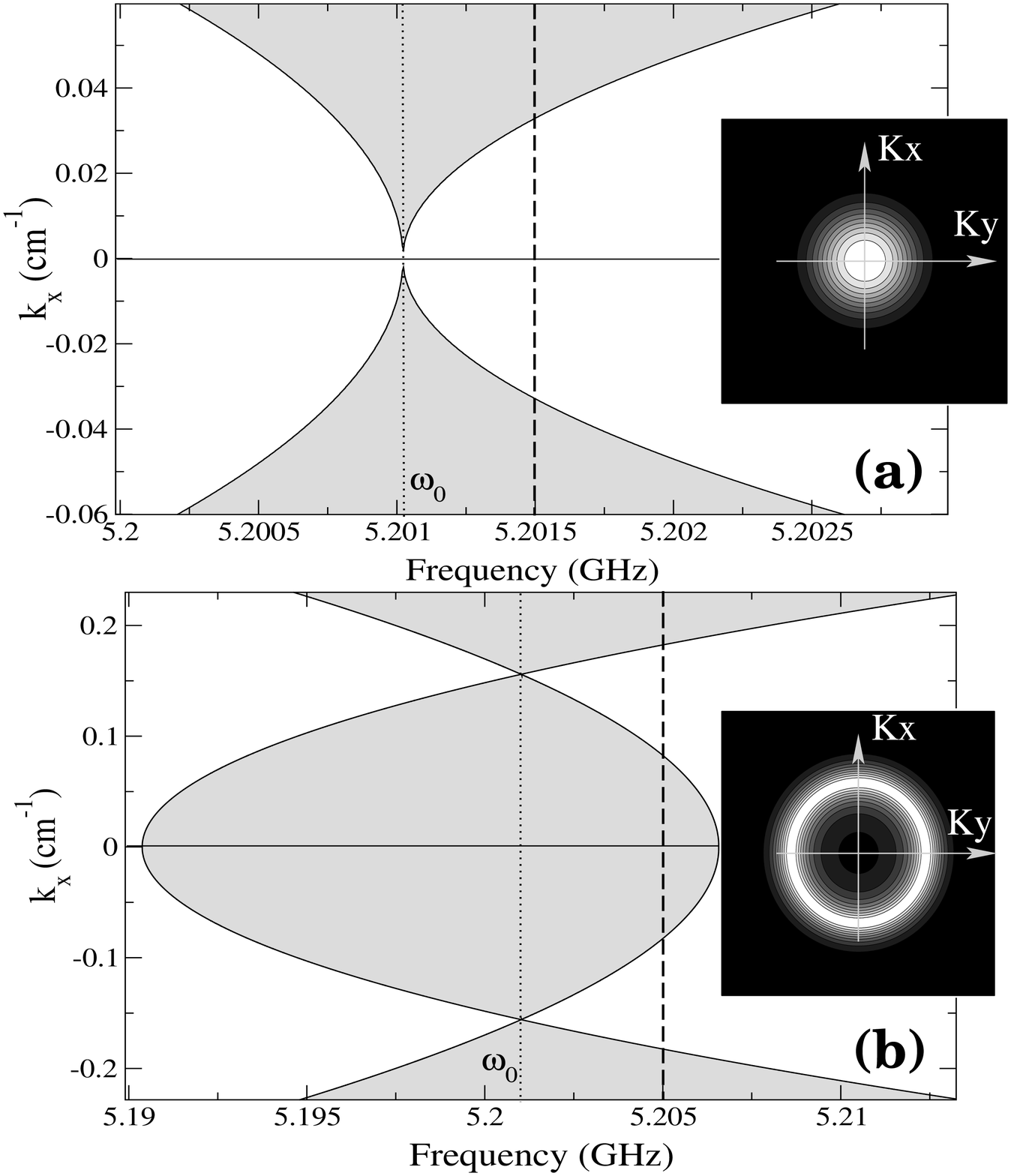}{gap_structure}{Band-gap structure on the
parameter plane $(\omega,k_x)$ with gaps shaded. (a) Transmission
band corresponds to a normal incidence; (b) Transmission band
corresponds to an oblique incidence. A dotted line is the
frequency $\omega_0$. Insets show the beam transmission
coefficients at the frequencies marked by dashed lines.}

\pict{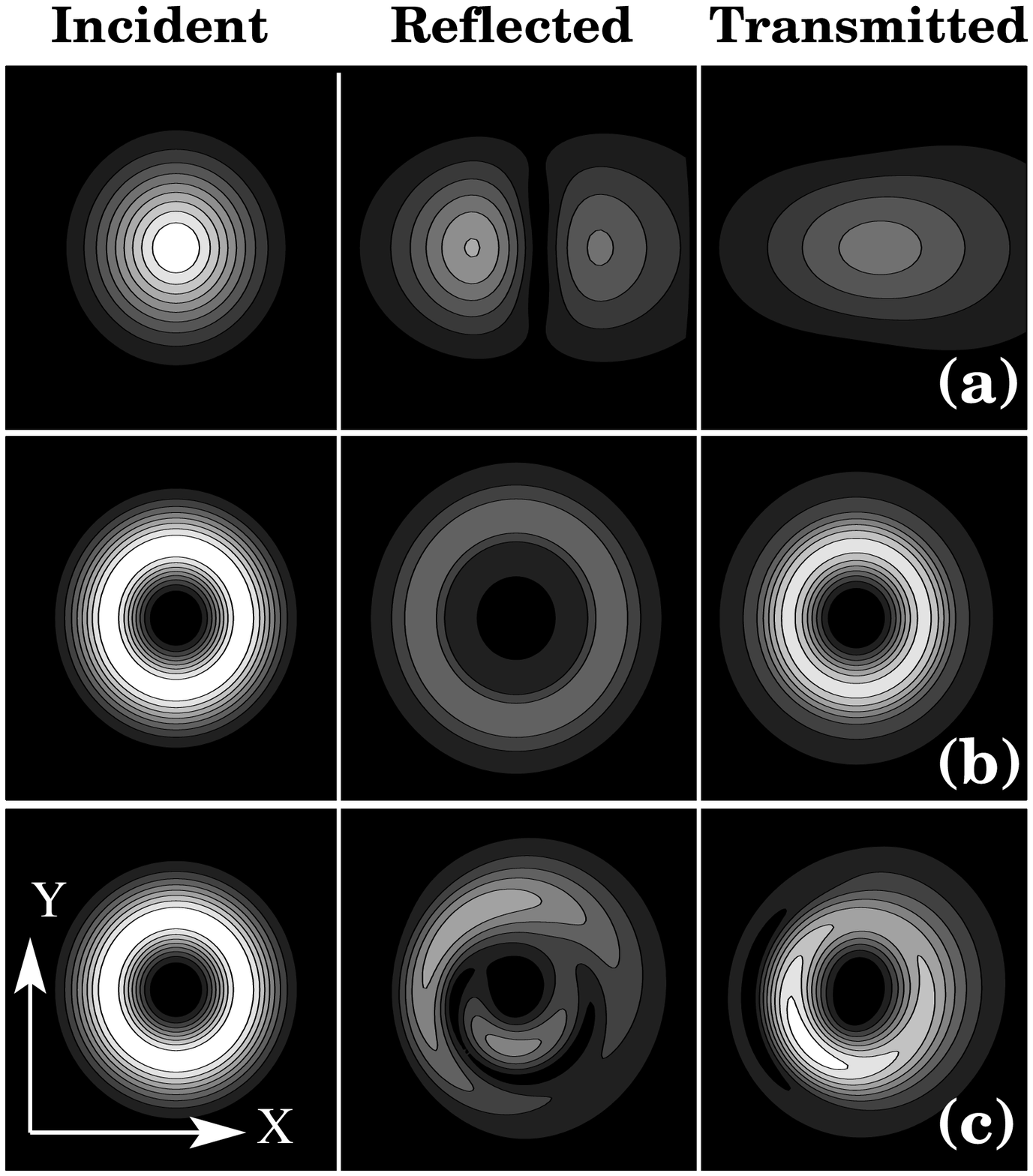}{scattering}{
Cross-section intensity profiles of the incident (left), reflected (middle), and transmitted (right) beams of various shapes: 
(a) Gaussian, (b-c) Gaussian with the vortex (topological) charge four at (b) the normal and (c) oblique incident angles.}

In comparison with the conventional case of the periodic structure
made of two different RH materials, the structure with LH layers
shown in Fig. \rpict{geom} possesses a distinctive bandgap of
a new type when $|k_{rz} a| = |k_{lz} b|$, the property recently
revealed in Refs. \cite{WuHe:arXiv, Chan:DownUnder}.
The inherent feature of the LH materials is their frequency
dispersion, so that in most of the cases we can find the
characteristic frequency $\omega_0$ such that $n_r = |n_l|$. Then,
a periodic structure consisting of the alternating LH and RH slabs of the same thickness ($a = b$) will formally possess a bandgap for all angles of incidence, with the only exception of the 100\% transmission resonances, which appear when the arguments of the sinus functions in Eq.~\reqt{TransferM} are equal to the whole multiple of $\pi$. Such highly unusual transmission can happen because the ``zero-index'' bandgap does not depend on the optical period of the structure, whereas usual Bragg-reflection and transmission resonances are highly sensitive to variations of the period and the incident angle.

It is important to study the transmission properties of a realistic system when the resonant conditions may not be exactly satisfied. To be specific, 
let us consider a periodic structure consisting of the
layers of LH metamaterial separated by air. We assume that the
composite material possesses the negative refractive index in the
microwave region with the effective dielectric permittivity and
magnetic permeability~\cite{Smith:2000-4184:PRL},
\begin{equation} \leqt{freqDisp}
   \epsilon(\omega) = 1 - \frac{\omega_p^2}{\omega^2} , \;\;\;\;\;
   \mu(\omega) = 1 - \frac{F\omega^2}{(\omega^2-\omega_r^2)} ,
\end{equation}
where $\omega_p/2\pi = 10$~GHz, $\omega_r/2\pi = 4$~GHz, and $F =
0.56$. The frequency range, where both $\epsilon$ and $\mu$ are
negative is (4 - 6)~GHz.
We plot the corresponding band-gap on the plane $(\omega, k_x)$ [see Fig.~\rpict{gap_structure}]. We reveal that for a fixed incident angle $(k_x)$ there appears a single frequency gap inside the transmission region. Quite a different situation occurs for the anglular dependence of the transmission coefficient at a given frequency, when there exists a narrow transmission band inside the bandgap. 

In Fig.~\rpict{gap_structure}(a), the period is chosen to satisfy
the condition $k_r a_0 = \pi$ (in our case, $a_0 = b_0 \approx
2.88$ cm), and the resonant transmission of electromagnetic waves
of the frequency $\omega_0$ occurs at the normal incidence. In
contrast, Fig.~\rpict{gap_structure}(b) corresponds to the case $a
= b = a_0+0.03$ \,cm when the transmission resonance occurs for
the waves at the incidence angle $\theta$: $k_r \cos{\theta}\,a =
\pi$. When the period is further increased,  multiple transmission
bands are observed. The width of the transmission band in the $ k
$-space can be controlled by tuning the frequency in the vicinity
of $\omega_0$. The insets in Fig.~\rpict{gap_structure} show the
cross-sections of the transmission coefficients for the
frequencies marked by dashed lines. Thus, one can make a periodic
structure with the transmission band at the desired angle of
incidence by a proper choice of the system period, while the width
of the transmission band can be adjusted by tuning the frequency.
Similarly, the structures with more complicated transmission
properties which include multiple rings in the transmission
coefficient {\em vs} wavevector can be obtained. Most importantly, we find that such narrow-band angular filters can have almost 100\% transparency in the transmission region. 

\pict{fig04.eps}{G_Hshift}{The Goos-H\"{a}nchen shift $\Delta$ of
the reflected beam {\em vs} $k_x$. 
Solid curves 1,2,3~--- numerically calculated shifts for the incident Gaussian beams 20, 60, and 100 cm wide, respectively. 
Dashed curve~--- the asymptotic result for wide beams.
Shaded area corresponds to the transmission band of the infinite periodic structure.}

We now use the standard transfer matrix method to analyse beam shaping by a finite composite structure consisting of 100 layers with the parameters
corresponding to the angular transmission coefficient shown in the
inset of Fig.~\rpict{gap_structure}(b). We consider the
transmission of (2+1)-dimensional beams in the paraxial
approximation when the angles of incidence are small and polarization effects are negligable. For a Gaussian beam incident at a normal angle, we observe almost complete reflection if the the angular spectrum width is comparable to the width of the transmission band. Transmission is only possible at the incident angles when the spectra overlap, and under such conditions the reflected beam has a {\em two-humped
shape}, see Fig.~\rpict{scattering}(a). On the other hand, vortex beams have a ring structure of the angular spectrum and can therefore be very effectively transmitted even at the normal incidence [Fig.~\rpict{scattering}(b)]. Similar to the case of a Gaussian beam, we observe that the weak reflected beam has a two-hump vortex structure. Small variations of the angle of incidence from the normal destroy the structure of the reflected and transmitted vortex beams and the results depend on the vortex topological charge [see Fig.~\rpict{scattering}(c)].

A beam reflected by an interface experiences a shift of its center relative to the center of the incident beam, and this classical effect is known as the Goos-H\"{a}nchen shift (see, e.g.~Ref.~\cite{Goos-Hanchen, Brekhovskikh:Waves:1960}). Reflection from a single LH slab results in a negative Goos-H\"{a}nchen shift~\cite{Lakhtakia:arXiv:0203084}, and it is interesting to calculate the beam shift for the layered structure.
The centers of the incident and reflected beams are found as
$    {\bf X}_{i,r} = \int_{-\infty}^{\infty} {\bf r_\bot} 
                            |E_{i,r}|^2 \, d{\bf r_\bot} /
        \int_{-\infty}^{\infty} | E_{i,r} |^2 \,d{\bf r_\bot}$,
where ${\bf r_\bot}$ is the transverse coordinate in the plane
$(x,y)$. For wide beams (i.e. the beams with a narrow spectrum),
the shift ${\bf \Delta} = {\bf X}_r - {\bf X}_i$ is determined by
the gradient of the phase $\phi$ of the reflection coefficient,
${\bf \Delta} = - \nabla_{{\bf k_\bot}} \, \phi$.
We calculate the Goos-H\"{a}nchen shift for the beam reflected by
the periodic RH/LH structure described above, using the parameters
corresponding to Fig.~\rpict{gap_structure}(b). With no loss of generality, we assume that the ($x,z$) plane is the plane of incidence, and therefore the beam is only shifted along the $x$ axis. We plot the shifts experienced by the Gaussian beam of various widths (see the curves 1-3 in Fig.~\rpict{G_Hshift}). We see that the shift of wide beams indeed approaches the the asymptotic result shown with dashed line in Fig.~\rpict{G_Hshift}.
The shift can be either positive and negative, and the absolute value increases dramatically for the angles of incidence close to the transmission band.

In conclusion, we have analyzed the transmission properties of a
one-dimensional photonic crystal composed of two materials with
positive and negative refractive indices. We have demonstrated unusual angular
dependencies of the beam transmission through such composite
structures which can be used for efficient beam reshaping. We have
calculated the Goos-H\"{a}nchen shift of the reflected beam which is shown to
increase dramatically for the angles of incidence close to the
transmission band.

\end{sloppy}

\begin{thebibliography}{99}

\bibitem{veselago}
V.G. Veselago, Sov. Phys. Uspekhi {\bf 10}, 509 (1968).

\bibitem{nature} A brief summary of the recent debates is presented in
L. Venema, Nature {\bf 420}, 119 (2002).

\bibitem{Pendry:2000-3966:PRL}
J.~B. Pendry, Phys. Rev. Lett. {\bf 85}, 3966 (2000).

\bibitem{Luo:2002-201104:PRB}
C. Luo, S.G. Johnson, J.D. Joannopoulos, and J.B. Pendry, Phys.
Rev. B {\bf 65}, 201104 (2002).

\bibitem{zhang} Z.M. Zhang and C.J. Fu, Appl. Phys. Lett. {\bf
80}, 1097 (2001).

\bibitem{gerardin} J. Gerardin and A. Lakhtakia, Micr. Opt. Tech.
Lett. {\bf 34}, 409 (2002). 

\bibitem{Chan:DownUnder}
C.T. Chan, Reported at "Photonic Crystals Down Under", 19-23
August, 2002 (Australian National University, Canberra, Australia)
(http://www.rsphysse.anu.edu.au/nonlinear/meeting/).

\bibitem{WuHe:arXiv}
L. Wu, S. He, and L. Chen, arXiv:physics/0211007 (2002).

\bibitem{Smith:2000-4184:PRL}
D.R. Smith, W. Padilla, D.C. Vier, S.C. Nemat-Nasser, and S.
Shultz, Phys. Rev. Lett. {\bf 84}, 4184 (2000).

\bibitem{Goos-Hanchen}
F. Goos and H. H\"{a}nchen, Ann. Phys. {\bf 2}, 87 (1947).

\bibitem{Brekhovskikh:Waves:1960}
L. Brekhovskikh, {\em Waves in Layered Media} (Academic Press, New York,
1960).

\bibitem{Lakhtakia:arXiv:0203084}
A. Lakhtakia, arXiv:physics/0203084 (2002).

\end{thebibliography}
\end{document}